\documentclass
[aps,prl,a4paper,superscriptaddress,amsmath,amssymb,twocolumn]{revtex4}%
\usepackage{amsfonts}
\usepackage{amsmath}
\usepackage{amssymb}
\usepackage{graphicx}%
\providecommand{\U}[1]{\protect\rule{.1in}{.1in}}
\begin{document}
\title{Efficient and accurate treatment of electron correlations from first-principles}

\author{Y. X. Yao}
\affiliation{Ames Laboratory--US DOE and Department of Physics and Astronomy, Iowa State
University, Ames, Iowa 50011, USA}
\author{J. Liu}
\affiliation{Ames Laboratory--US DOE and Department of Physics and Astronomy, Iowa State
University, Ames, Iowa 50011, USA}
\author{C. Liu}
\affiliation{Ames Laboratory--US DOE and Department of Physics and Astronomy, Iowa State
University, Ames, Iowa 50011, USA}
\author{W. C. Lu}
\affiliation{State Key Laboratory of Theoretical and Computational Chemistry, Institute of
Theoretical Chemistry, Jilin University, Changchun, Jilin 130021, China }
\affiliation{College of Physical Science and Laboratory of Fiber Materials and Modern
Textile, Growing Base for State Key Laboratory, Qingdao University, Qingdao,
Shandong 266071, China}
\author{C. Z. Wang}
\affiliation{Ames Laboratory--US DOE and Department of Physics and Astronomy, Iowa State
University, Ames, Iowa 50011, USA}
\author{K. M. Ho}
\affiliation{Ames Laboratory--US DOE and Department of Physics and Astronomy, Iowa State
University, Ames, Iowa 50011, USA}

\begin{abstract}
We present an efficient \textit{ab initio} method for calculating the
electronic structure and total energy of strongly correlated electron systems.
The method extends the traditional Gutzwiller approximation for one-particle
operators to the evaluation of the expectation values of two particle
operators in a full many-electron Hamiltonian. The method is free of
adjustable Coulomb parameters, and has no double counting issues in the
calculation of total energy, and has the correct atomic limit. We demonstrate
that the method describes well the bonding and dissociation behaviors of the
hydrogen and nitrogen clusters. We also show that the method can
satisfactorily tackle great challenging problems faced by the density
functional theory recently discussed in the literature. The computational
workload of our method is similar to the Hartree-Fock approach while the
results are comparable to high-level quantum chemistry calculations.

\end{abstract}
\maketitle

It is one of the outstanding challenges in physics, chemistry, and materials
science to develop robust and efficient theoretical and computational methods
to accurately calculate the electronic structure and total energy of materials
containing strongly correlated electrons ~\cite{nota-DOE-nuclear}. While
accurate methods are available from quantum chemistry approaches (e.g.,
configuration interaction (CI)), these methods are too expensive for condensed
matter systems. On the other hand, density functional theory (DFT) and related
computational codes based on the Kohn-Sham approach~\cite{DFTHK,DFTKS} have
been well developed, and are highly effective and successful for predicting
the structures and properties of many materials, but they fail for systems
with strongly correlated electrons. In the last three decades, there have been
intensive efforts in developing new approaches to solve the outstanding
problems in correlated electron systems
~\cite{LDA+U91,LDA+U,DMFT,DMFTPu,LDA+U+DMFT,DMFT_Chan,DMFT_CHEM_Reichman,
DMET,HoGDFT,YaoMBS,YaoH,LDA+G0,XDaiGAB,Guangtao10,Gmethod,Schickling12,Cerium}. Among
these new developments, local-density approximation plus on-site Coulomb
interaction parameter U (LDA+U)~\cite{LDA+U91,LDA+U}, LDA+dynamical mean field
theory~\cite{DMFT,DMFTPu,LDA+U+DMFT}, and
LDA+Gutzwiller~\cite{HoGDFT,YaoMBS,LDA+G0,XDaiGAB,Guangtao10,Gmethod,Schickling12,Cerium}
have emerged as the most popular methods for treating strongly-correlated
electrons in solid-state systems. These methods handle electron correlations
through the adjustable on-site Coulomb interaction parameters, while keeping
the full description of the electronic structure through LDA. From the aspect
of the theoretical predictive power, it is highly desirable to have a fully
self-consistent \textit{ab initio} theory that can treat correlated electron
systems without adjustable parameters and with computational speed comparable
to LDA or Hartree-Fock (HF) calculations~\cite{CarNSO}.

In this letter, we present an \textit{ab initio} method for the electronic
structure and total energy calculations of strongly correlated electron
systems without adjustable Coulomb parameters. In our approach, the
commonly-adopted Gutzwiller approximation (GA) for evaluating the one particle
density matrix~\cite{GA1B,KRSB,BuneMBGA,GASB} is extended to treat the
evaluation of the two-particle correlation matrix of the system. This
approximation, which we call the correlation matrix renormalization (CMR)
approximation~\cite{CMRPRB}, allows the expectation value of a many-electron
Hamiltonian with respect to Gutzwiller variational wave function (GWF) to be
evaluated with reduced computational complexity. We show that the method
describes well the bonding and dissociation behaviors of hydrogen and nitrogen
clusters in comparison with the accurate and expensive quantum chemistry
calculations. Furthermore, some of the most challenging problems faced by
Kohn-Sham DFT-based calculations recently discussed in the literature
~\cite{CohenHZ,CohenHZTalk} can also be readily solved by our method. The
method has no double counting issues in the calculation of total energy, and
produces the correct atomic limit. The computational workload scales as
N$^{4}$ or better for systems of size N, similar to the HF
calculations\cite{CMRPRB}.

We start with the full \textit{ab initio} Hamiltonian for an interacting
many-electron system in the second quantization form
\begin{align}
\mathcal{H} &  =\sum_{i\Gamma}E_{i\Gamma}\left\vert \Gamma_{i}\right\rangle
\left\langle \Gamma_{i}\right\vert +\sum_{i\alpha j\beta\sigma}^{\prime
}t_{i\alpha j\beta}c_{i\alpha\sigma}^{\dagger}c_{j\beta\sigma}\nonumber\\
&  +\frac{1}{2}\sum_{i\alpha j\beta k\gamma l\delta\sigma\sigma^{\prime}%
}^{\prime}U_{i\alpha j\beta}^{k\gamma l\delta}c_{i\alpha\sigma}^{\dagger
}c_{j\beta\sigma^{\prime}}^{\dagger}c_{k\gamma\sigma^{\prime}}c_{l\delta
\sigma}\label{H}%
\end{align}
Here $i$, $j$, $k$, $l$ are the atomic site indices. $\alpha$, $\beta$,
$\gamma$, $\delta$ are orbital indices and $\sigma$ is the spin index.
$\{\Gamma_{i}\}$ are eigenstates of the local many-body Hamiltonian. The first
term is the local on-site part which has been singled out for exact treatment,
$E_{i\Gamma}$ is the energy of the local many-electron configuration
$\left\vert \Gamma_{i}\right\rangle $. The second and third terms describe the
non-local one-body and two-body contributions. All interactions are included
in this Hamiltonian without any adjustable parameters. When evaluating this
Hamiltonian with the full CI wave function, one obtains an exact expression of
the total energy which consists of non-local one-particle and two-particle
density matrices in addition to the local on-site contributions. In our CMR
approach, we evaluate the Hamiltonian in Eq. \ref{H} with the GWF of the form%
\begin{equation}
\left\vert \Psi_{GWF}\right\rangle =%
%TCIMACRO{\dprod \limits_{i}}%
%BeginExpansion
{\displaystyle\prod\limits_{i}}
%EndExpansion
\left(  \sum_{\Gamma}g_{i\Gamma}\left\vert \Gamma_{i}\right\rangle
\left\langle \Gamma_{i}\right\vert \right)  \left\vert \Psi_{0}\right\rangle
\label{GWF}%
\end{equation}
Where $\left\vert \Psi_{0}\right\rangle $ is a non-interacting electron
wavefunction, i.e., a single Slater determinant. $g_{i\Gamma}$ is the
variational parameter which determines the occupation probability of the
on-site configuration $\left\vert \Gamma_{i}\right\rangle $. The central part
of the Gutzwiller approach is the suppression of the energetically unfavorable
atomic configurations in the many-body wave function. Using the GWF of Eq.
\ref{GWF} and adopting the generally accepted GA for the expectation value of
a one-particle operator \cite{BuneMBGA,GASB}, the total energy of the system
in our CMR scheme can be expressed as\begin{widetext}
\begin{align}
E &  =\sum_{i\Gamma}E_{i\Gamma}p_{i\Gamma}+\sum_{i\alpha j\beta\sigma}%
^{\prime}t_{i\alpha j\beta}z_{i\alpha\sigma}^{j\beta}\left\langle
c_{i\alpha\sigma}^{\dagger}c_{j\beta\sigma}\right\rangle _{0}\nonumber\\
&  +\frac{1}{2}\sum_{i\alpha j\beta k\gamma l\beta\sigma\sigma^{\prime}%
}^{\prime}U_{i\alpha j\beta}^{k\gamma l\delta}\left(  z_{i\alpha\sigma
}^{l\delta}\left\langle c_{i\alpha\sigma}^{\dagger}c_{l\delta\sigma
}\right\rangle _{0}z_{j\beta\sigma^{\prime}}^{k\gamma}\left\langle
c_{j\beta\sigma^{\prime}}^{\dagger}c_{k\gamma\sigma^{\prime}}\right\rangle
_{0}-\delta_{\sigma\sigma^{\prime}}z_{i\alpha\sigma}^{k\gamma}\left\langle
c_{i\alpha\sigma}^{\dagger}c_{k\gamma\sigma}\right\rangle _{0}z_{j\beta\sigma
}^{l\delta}\left\langle c_{j\beta\sigma}^{\dagger}c_{l\delta\sigma
}\right\rangle _{0}\right)  \nonumber\\
&  +E_{c}\label{E}%
\end{align}
\end{widetext}where $P_{i\Gamma}$ is the occupation weight of the
configuration $\left\vert \Gamma_{i}\right\rangle $. $z_{i\alpha\sigma
}^{j\beta}$\ can be evaluated following the standard GA rule for one-particle
hopping operators, i.e., $z_{i\alpha\sigma}^{j\beta}=z_{i\alpha\sigma
}z_{j\beta\sigma}$ if $\left(  i\alpha\right)  \neq\left(  j\beta\right)  $
and $1$ otherwise. To reach the expression Eq. \ref{E}, the validity of Wick's
theorem has been assumed. $E_{c}$ is the residual correlation energy due to
the approximation involved in the CMR approach. In general, $E_{c}$ can be
determined by comparing the total energies from the CMR with that from
accurate CI or quantum Monte Carlo calculations for some exactly solvable
structures. Since the dominant local onsite electron correlation effect has
been taken into account by the GWF, the residual correlation energy due to the
CMR approximation is expected to be small. In the test cases to be shown in
this letter, we find that one way to include the effects of $E_{c}$ is to
modify the renormalization $z$-factor obtained from the GA.

\begin{figure*}[th]
\centering
\includegraphics[
width=5.5in
]{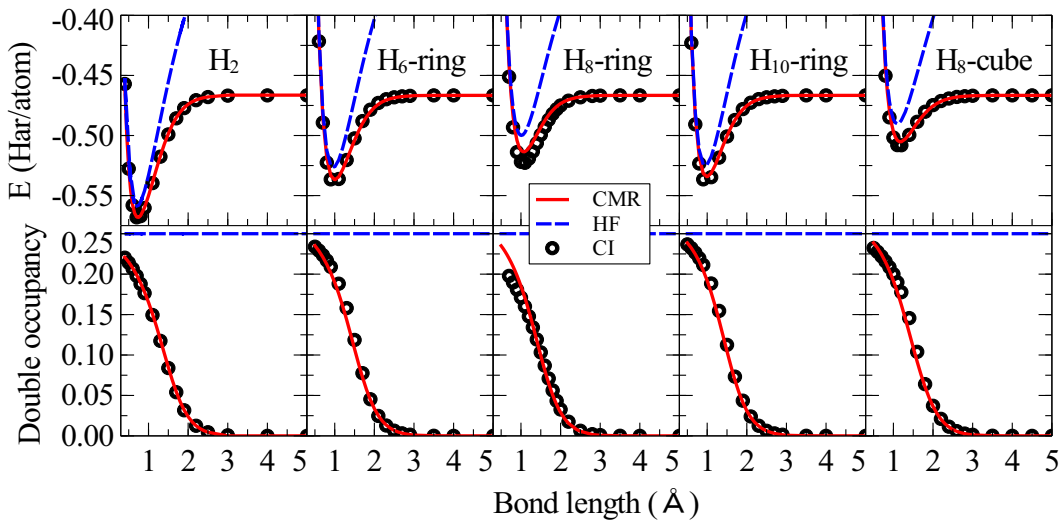}\caption{(Color online) The total energy (upper panel) and double
occupancy weight (lower panel) of H$_{n}$ clusters as a function of bond
length calculated from the CMR method, which agree very well with the results
from the exact CI calculations. The CMR results are also much better than the
HF results. The calculations are done using a minimal basis set.}%
\label{hsto}%
\end{figure*}

We first demonstrate the CMR method by studying the dissociation behavior of
the hydrogen molecules. The dissociation behavior of these hydrogen molecules
has been the testing ground of methods for correlated electron calculations,
because the electron correlation changes from the weak to strong regime as the
hydrogen bond length increases. For these systems, the residual correlation
energy is included by replacing the renormalization $z$-factor obtained using
the density-density type GA \cite{BuneMBGA} by the functional of $z$, i.e.,
$f(z)$. The $f(z)$ is determined by requiring that the total energy (without
explicit $E_{c}$ term) and the probability of the local double occupancy for a
H$_{2}$ dimer obtained from the CMR calculation to be the same as the exact CI
results. We first performed the test using the minimal basis set (one $s$
orbital for each H atom). In this case, the total energy and double occupancy
probability from the CI calculation can be solved analytically and the $f(z)$
has an analytical form in term of $z$. Using the $f(z)$ analytically
determined from the reference system H$_{2}$, CMR calculations are performed
for H$_{6}$-ring, H$_{8}$-ring, H$_{10}$-ring and H$_{8}$-cube structures. The
results from our CMR calculations are presented in Fig. \ref{hsto} in
comparison with the full CI or the highly accurate multi-configurational
self-consistent field (MCSCF) results. We found the bonding and dissociation
behavior of the hydrogen clusters calculated from the CMR method agrees very
well with the result from the high level quantum chemistry calculations. In
contrast, the HF\ results show large systematic errors, especially at large
separations where the electron correlation effect becomes prominent, as
evidenced by the strong suppression of the energetically unfavorable local
electron double occupancy weight.

\begin{figure}[pb]
\centering\includegraphics[
width=3.0in
]{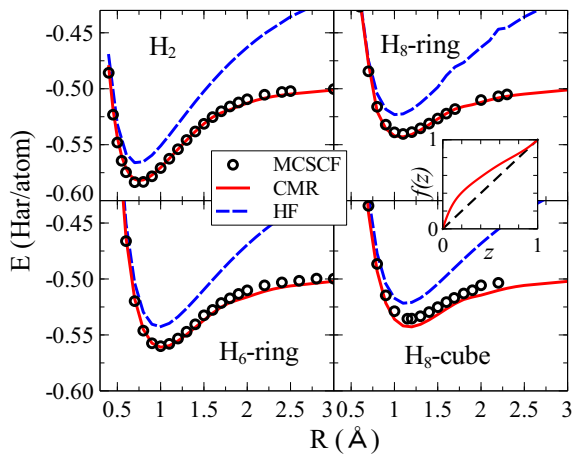}\caption{(Color online) The total energy of hydrogen clusters as a
function of bond length calculated from the CMR method, which agree well with
the results from the high-level quantum chemistry CI or MCSCF calculations.
The HF results are also shown for comparison. The calculations are done using
large basis set. Insets: $f(z)$ obtained by fitting the CMR total energy and
local configuration weights with the exact full CI results.}%
\label{Hlb}%
\end{figure}

We further tested the CMR method for the dissociation behavior of hydrogen
clusters using a large basis set of 6-311G**, which contains 3 $s$-orbitals
plus 3 $p$-orbitals. In this case, the $f(z)$ needs to be determined
numerically by fitting the CMR energies and the local configuration occupation
probabilities of the H$_{2}$ dimer to the exact CI results. Using such a
numerically constructed functional $f(z)$, we have performed the CMR
calculations for H$_{6}$-ring, H$_{8}$-ring, and H$_{8}$-cube with the same
large basis set. In Fig. \ref{Hlb}\ we show that the CMR\ method yields again
very good bonding and dissociation curves in close agreement with the MCSCF
calculations. The inset of Fig. \ref{Hlb} shows the behavior of $f(z)$, which
scales like $\sqrt{z}$ at small $z$ and approaches $z$ as $z$ goes to $1$.

\begin{figure}[ptb]
\centering
\includegraphics[
width=3.3in
]{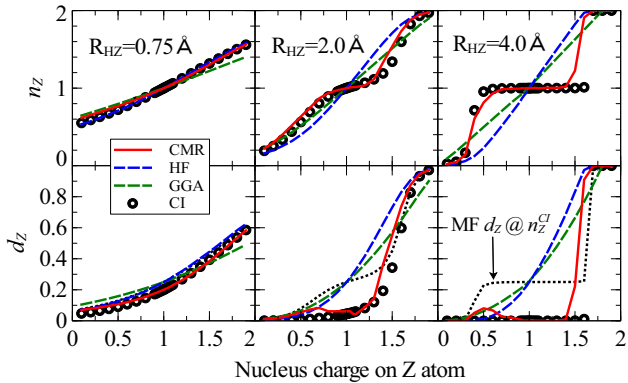}\caption{ (Color online) The electron occupation number $n_{Z}$
(upper panel) and the local double occupancy weight $d_{Z}$\ (lower panel)\ as
a function of the nucleus charge on the Z atom calculated with the CMR, HF,
GGA and CI methods for the HZ dimer at three seperations: near equilibrium
(left), close to bond breaking (middle) and beyond (right). The dotted line in
the lower panel shows the mean field double occupancy weight evaluated at the
CI electron occupation of the Z atom. }%
\label{hz}%
\end{figure}

Very recently, Cohen, et al used some prototype systems to show the dramatic
errors in the DFT-based calculations. These errors stem from the fact that the
current approximations used in DFT calculations miss the energy derivative
discontinuity with respect to the total electron
number\cite{DerivativeDisE,CohenHZ,CohenHZTalk}. The prototype systems to
reveal the failure of DFT are the stretched few-electron systems, e.g.,
one-electron systems like HZ$^{\left\{  1e\right\}  }$ and HZH$^{\left\{
1e\right\}  }$ and two-electron systems like HZ$^{\left\{  2e\right\}  }$ with
Z being the proton with nucleus charge $Z$ varying between 0 and 2. While the
electron density from the exact calculations shows dramatic discontinuous
changes in real space with a slight variation of $Z$ near some critical points
at large separations, all the DFT calculations predict an artificial
continuous variation of the electron density \cite{CohenHZ}. Remarkably, our
CMR method gives exact solutions for any single-electron systems, as it can be
easily proved that the orbital renormalization factors are constantly one and
the method reproduces the HF results, which are exact in the special class of
one-electron systems. It is also remarkable that our CMR method yields the
exact bonding and dissociation behaviors for both H$_{2}^{+}$ and H$_{2}$ (see
Fig. \ref{Hlb}), while all the available DFT calculations can hardly describe
both cases equally well~\cite{CohenHZTalk}. One can further show that because
the CMR method reaches the correct atomic solutions at the large separation
limit, the exact discontinuous electron transfer observed in the HZ$^{\left\{
2e\right\}  }$ system at large separations can be well reproduced. In Fig.
\ref{hz} we compare the electron occupation and double occupancy weight of Z
atom from the CMR, HF, DFT with the generalized gradient approximation (GGA)
and CI. Although all the methods predict similar results near equilibrium bond
length ($\sim$0.75\r{A}), the CMR method shows significant improvements over
the mean field HF and GGA and follows closely the exact CI\ results with
increasing separations, even at the chemically crucial bond breaking region
($\sim$2\r{A}) and beyond. The underlying physics for the large errors of the
simple mean field approaches like the HF and GGA can be understood by noting
that the mean field double occupancy weight evaluated using the CI\ orbital
occupation, shown as the dotted line in the lower panel of Fig. \ref{hz}, can
severely deviate from the exact CI double occupancy weight---manifesting the
multi-configuration nature of the exact solution which is beyond the single
Slater determinant description.

\begin{figure}[ptb]
\centering
\includegraphics[
width=3.3in
]{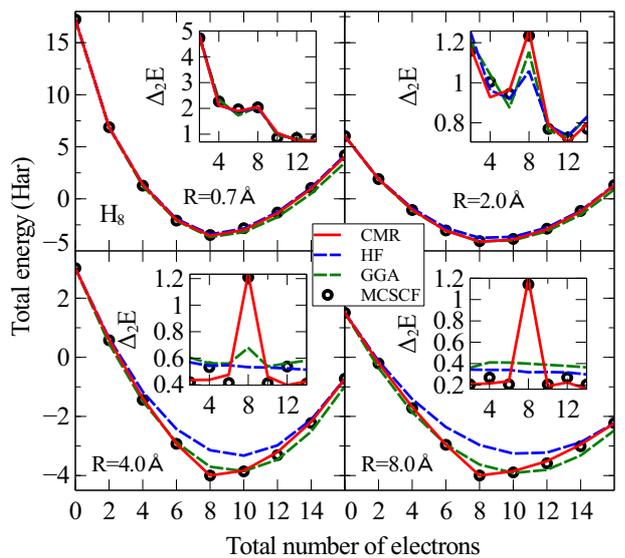}\caption{ (Color online) The total energy $E$ as a function of total
number of electrons $N_{e}$ obtained from the CMR, HF, GGA and MCSCF
calculations for the H$_{8}$ cube with increasing bond length $R$. Insets: the
corresponding second order energy difference $\Delta_{2}E$.}%
\label{h8}%
\end{figure}

Another challenging prototype system is the H$_{8}$ cluster with varying
electron filling \cite{CohenHZTalk}. The exact solution predicts a relatively
big energy gap for the system at large separations and half-filling $N_{e}=8$;
while all the DFT calculations fail to reproduce this result because of the
incapability to treat the strong electron correlation effects. In Fig.
\ref{h8} we show the total energy of the H$_{8}$ cube from the CMR, HF, GGA
and MCSCF calculations as a function of even number of electron filling, which
keeps the system to have the closed shell ground state solution
\cite{CohenHZTalk}. While all the four theories give similar total energies at
the small bond length, the discrepancy between them becomes increasingly large
with expanding the H$_{8}$ cluster. Remarkably, the CMR energies agree with
the highly accurate MCSCF results very well for all the bond separations and
electron fillings, which proves that the key many-body correlation physics in
this system has been perfectly captured by the CMR method. A better comparison
between the four levels of theories is presented by the energy gap, defined as
the second order finite difference $\Delta_{2}E=E\left(  N_{e}+2\right)
+E\left(  N_{e}-2\right)  -2E\left(  N_{e}\right)  $, as shown in the insets
of Fig. \ref{h8}. Clearly, as the bond length increases, or the electron
correlation effects become stronger, the simple mean field HF and GGA energy
gap shows larger deviations from the exact gap, especially the gap at half
filling. In contrast, the CMR calculations yield energy gaps in excellent
agreement with the MCSCF results in all the cases.

\begin{figure}[ptb]
\centering\includegraphics[
width=3.0in
]{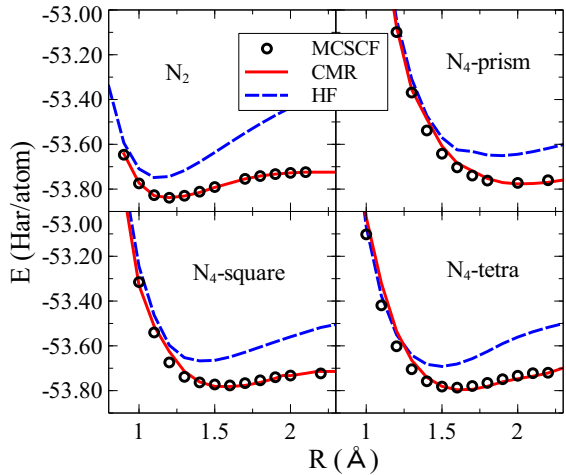}\caption{(Color online) The total energy of the nitrogen clusters
as a function of bond length calculated from the CMR method, which agrees well
with the results from the high-level CI or MCSCF calculations. The HF results
are also shown for comparison.}%
\label{Nsto}%
\end{figure}

The CMR method is also successfully applied to systems with atoms containing
multiple correlated orbitals, e.g., nitrogen clusters. For computational
convenience, we describe the nitrogen atom with the minimum basis set and
choose the $2s$ and $2p$ as the correlated orbitals. The same idea can be
equally well carried over to the large basis calculations as shown previously
for the hydrogen clusters. Two functionals, $f_{s}\left(  z_{s}\right)  $ and
$f_{p}\left(  z_{p}\right)  $, are introduced to modify the renormalization
coefficients of $2s$ and $2p$ orbitals. The specific functional forms,
following the procedure in the calculations of hydrogen clusters, are
determined by matching the CMR total energy, $E$, and local correlated Fock
state occupation probabilities, $p_{\Gamma}$, with the exact CI results of the
N$_{2}$ dimer. We apply the method to calculate binding energy curves of three
nitrogen clusters of different geometries, i.e., the square, diamond and
tetragonal shapes. In Fig.~\ref{Nsto} we show the total energy as a function
of bond length from the CMR, HF\ and MCSCF calculations. The good agreement
between the CMR and MCSCF energies for all the structures demonstrates the
good transferability of our method.

In summary, we have developed an efficient method for calculating the
electronic structure and total energy of the systems with strong electron
correlations. The method is based on the Gutzwiller type variational
wavefunction and adopts a correlation matrix renormalization approximation in
which both one-particle density and two-particle correlation matrices at mean
field level are renormalized according to the local electron correlation
effects. While the computation workload of this new approach is similar to
that in HF calculations, the calculation results are much more accurate. The
benchmark results for the bonding and dissociation behaviors of the hydrogen
and nitrogen clusters show that our method well reproduces the results from
the accurate and yet expensive quantum chemistry CI and MCSCF calculations.
The CMR method is also demonstrated to be accurate for treating the electron
correlation effects in some prototype systems where the current DFT and HF
calculations fail. The extension of the method to crystalline solids is
straightforward and promising. The work along this direction is underway.

\begin{acknowledgments}
This work was supported by the U. S. Department of Energy, Basic Energy
Sciences, Division of Materials Science and Engineering under the Contract No.
DE-AC02-07CH11358 including the computer time support from the National Energy
Research Supercomputing Center (NERSC) in Berkeley, CA.
\end{acknowledgments}

\bibliographystyle{apsrev}
\bibliography{REF}

\end{document}